
\magnification=\magstep1

\baselineskip=14pt
\overfullrule0pt
\nopagenumbers
\newcount\knum
\knum=1
\newcount\uknum
\uknum=1
\newcount\znum
\znum=1
\newcount\notenumber

\def\Za{\the\uknum.\the\znum \global\advance\znum by 1}
\def\ukneu{\znum=1\global\advance\uknum by 1}
\font\tbfontt=cmbx10 scaled \magstep0
\font\gross=cmbx10 scaled \magstep2
\font\mittel=cmbx10 scaled \magstep1
\font\ittel=cmr10 scaled \magstep1

\font\TT=cmcsc10 scaled \magstep0

\font\eightrm=cmr8

\font\eightit=cmti8 scaled \magstep0
\def\sqr#1#2{{\vcenter{\vbox{\hrule height.#2pt\hbox{\vrule width.#2pt
height#1pt \kern#1pt \vrule width.#2pt}\hrule height.#2pt}}}}
\def\square{\mathchoice\sqr34\sqr34\sqr{2.1}9\sqr{1.5}9}
\def\QED{\smallskip\rightline{$\square$ \quad\qquad\null}}
\def\Definition{\vskip 0.3 true cm\sl {\TT Definition \Za.}{ }}
\def\Theorem{\vskip 0.3 true cm\sl {\TT Theorem \Za.}{ }}

\def\Remark{\vskip 0.3 true cm\sl {\TT Remark \Za.}{ }\rm}

\def\Corollary{\vskip 0.3 true cm\sl {\TT Corollary \Za.}{ }}
\def\Lemma{\vskip 0.3 true cm\sl {\TT Lemma \Za.}{ }}
\def\Proof{\TT Proof:{ }\rm}

\def\cz{{\rm C\hskip-4.8pt\vrule height5.8pt\hskip5.3pt}}

\def\zz{{\rm Z\hskip -4pt Z}}
\def\eins{{\rm 1\hskip -2pt I}}

\def\mapright#1{\smash{\mathop{\longrightarrow}\limits^{#1}}}

\def\mapdown#1{\Big\downarrow\rlap{$\vcenter{\hbox{$\scriptstyle#1$}}$}}

\def\mapup#1{\uparrow\rlap{$\vcenter{\hbox{$\scriptstyle#1$}}$}}

\def\lim{\mathop{\rm lim}}
\def\hir{\mathop{\rightleftharpoons}}

\def\End{{\hbox{\rm End}}}
\def\ker{{\hbox{\rm ker}}}

\def\Hom{{\hbox{\rm Hom}}}

\parindent=0pt
\font\eightrm                = cmr8
\font\eightsl                = cmsl8
\font\eightsy                = cmsy8
\font\eightit                = cmti8

\font\eighti                 = cmmi8
\font\eightbf                = cmbx8
\def\petit{\def\rm{\fam0\eightrm}
\textfont0=\eightrm 
 \textfont1=\eighti 
 \textfont2=\eightsy 
 \def\it{\fam\itfam\eightit}
 \textfont\itfam=\eightit
 \def\sl{\fam\slfam\eightsl}
 \textfont\slfam=\eightsl
 \def\bf{\fam\bffam\eightbf}
 \textfont\bffam=\eightbf 
 \normalbaselineskip=9pt
 \setbox\strutbox=\hbox{\vrule height7pt depth2pt width0pt}
 \normalbaselines\rm}
\newdimen\refindent
\def\begref{\vskip1cm\bgroup\petit
\setbox0=\hbox{[Bi,Sc,So]o }\refindent=\wd0
\let\sl=\rm\let\INS=N}

\def\ref#1{\filbreak\if N\INS\let\INS=Y\vbox{\noindent\tbfontt
Literatur\vskip1cm}\fi\hangindent\refindent
\hangafter=1\noindent\hbox to\refindent{#1\hfil}\ignorespaces}

\long\def\fussnote#1#2{{\baselineskip=9pt
\setbox\strutbox=\hbox{\vrule height 7pt depth 2pt width 0pt}%
\petit\noindent\footnote{\noindent #1}{#2}}}
\rightline{hep-th/9503153}
\rightline{CPT-95/P.3166}
\rightline{Mannheimer Manuskripte 186}
\rightline{January 1995}
\vskip 4cm
\phantom{prelim.version}
\centerline{\gross The generalized Lichnerowicz formula}
\smallskip
\centerline{\gross and analysis of Dirac operators }
\vskip 1cm

\centerline{\ittel Thomas Ackermann\footnote{$^1$}{\eightrm
e-mail: ackerm@euler.math.uni-mannheim.de}\footnote{$^{\;\dag}$}{\eightrm
Address after April 1, 1995: Wasserwerkstr. 37, 68309 Mannheim, F.R.G.}
and
J\"urgen Tolksdorf\footnote{$^2$}{\eightrm
e-mail: tolkdorf@cptsu4.univ-mrs.fr}\footnote{$^{\;\ddag}$}{\eightrm Supported
by the European Communities,
contract no.\hskip -0.3cm \hbox{\petit\noindent\rm ERB 401GT 930224}}
}

\vskip 0.5cm
\centerline{\vbox{\hsize=4.7 true in \rm\noindent
$^1$\ Fakult\"at f\"ur Mathematik \& Informatik, Universit\"at
Mannheim \break \phantom{$^1$} D-68159 Mannheim, F.R.G.}}
\smallskip
\centerline{\vbox{\hsize=4.7 true in \rm\noindent
$^2$\ Centre de Physique Th${\acute {\rm e}}$orique, CNRS Luminy,
Case 907 \phantom{aaaaaa} \break
\phantom{$^2$} F-13288 Marseille Cedex 9, France }}
\vskip 1.1cm
\centerline{\vbox{\hsize=5.0 true in \petit\noindent
\bf Abstract. \rm We study Dirac operators acting on sections of a
Clifford module ${\cal E}$\ over a Riemannian manifold $M$.
We prove the intrinsic decomposition formula for their square, which is
the generalisation of the well-known formula due to Lichnerowicz [L].
This formula enables us to distinguish
Dirac operators of simple type. For each Dirac operator of this natural
class the local Atiyah-Singer index theorem holds. Furthermore, if
$M$\ is compact and ${{\petit \rm dim}\;M=2n\ge 4}$, we derive an
expression for the Wodzicki function $W_{\cal E}$, which is defined
via the non-commutative residue on the space of all Dirac operators
${\cal D}({\cal E})$. We calculate this function for certain
Dirac operators explicitly. From a physical point of view this
provides a method to derive gravity, resp. combined gravity/Yang-Mills
actions from the Dirac operators in question.}}
\rm \phantom{1991}
1991 AMS: \it 53A50, 58G10, 58G15, 81E13
\vfil\break  \rm
\advance\hsize by -0.5true in
\advance\vsize by -0.9true in
\advance\hoffset by 0.45in
\advance\voffset by 0.7in
\pageno=1
\topskip=1.2cm plus 0.2cm
\parindent=0pt

{\mittel 1. Introduction}
\vskip 0.7cm
In 1928, when develloping the quantum theory of
the electron, Dirac introduced his famous first-order operator -
the square-root of the so-called wave-operator (d'Alembertian operator).
Generalisations of this operator, called `Dirac operators',
have come to play a fundamental role in the mathematics of our
century, particularly in the interrelations between topology, geometry
and analysis. To mention only a few applications, Dirac operators are
the most
important tools
in proving the general Atiyah-Singer index theorem for
pseudo-differential operators both
in a topological [AS] and more analytical way  [G].
Dirac operators also assume a significant place in Connes'
non-commutative geometry [C] as the main ingredient of a K-cycle,
where they encode the geometric structure of the underlying
non-commutative `quantum-spaces'.
In mathematical physics,
in contrast,
Dirac operators had almost fallen into
oblivion,
except in modifications of Dirac's original
application and
Witten's proof of the positive mass conjecture in general
relativity [W] where he used the Dirac operator on the spinor
bundle $S$\ of a four-dimensional Lorentzian manifold.
More recently, however, Connes and Lott derived
the action
of the standard model of elementary particles using a special
K-cycle [CL]. Moreover,
the Einstein-Hilbert gravity action can be reproduced via
the non-commutative residue
from a Dirac operator
on a Clifford module ${\cal E}$\
over a four-dimensional Riemannian manifold as it was
explicitly shown in [K] and [KW].
\smallskip
Nowadays a Dirac operator
on a Clifford module ${{\cal E}={\cal E}^+\oplus {\cal E}^-}$\
is understood as an odd-parity first order operator $D\colon \Gamma({\cal
E}^\pm)
 \rightarrow
\Gamma({\cal E}^\mp)$, whose square $D^2$\ is a generalized laplacian.
In view of this rather general definition,
all Dirac operators of the examples mentioned above
are in a sense structurally the most simple ones as
they correspond to Clifford connections on the respective Clifford
module
${\cal E}$. Hence, it may not be surprising that other types of
Dirac operators have more or less failed to be studied
in the literature.
\smallskip
Here our subject is a more thorough treatment of
Dirac operators acting on sections of a Clifford module ${\cal E}$\
under two aspects:
We prove the intrinsic decomposition formula (3.14) for their
square, which is the generalisation of the well-known formula
due to Lichnerowicz [L], and apply this formula to calculate
explicitly the
the Wodzicki function $W_{\cal E}$\
on the space of all Dirac operators ${\cal D}({\cal E})$\ in
theorem 6.4. This complex-valued function
$W_{\cal E}\colon {\cal D}({\cal E})\rightarrow \cz$\ is defined via the
non-commutative residue, which has been extensively studied
by Guillemin and Wodzicki (cf. [Gu], [W1], [W2]) and represents a
link between Dirac operators and (gravity-) action functionals as
already mentioned.
\smallskip
We now give a summary of the different sections:
After fixing our notation
we recall Quillen's statement (cf. [BGV]),
that `Dirac operators are a quantisation of the theory of connections'
in the first section.
More precisely, we show in lemma 2.1, that given any Dirac operator
$D$, there exists a connection $\nabla$\ on the Clifford module
${\cal E}$\ such that $D=c\circ \nabla$.
This is the essential property to prove
the generalized Lichnerowicz formula (3.14) in section 3 . Since
there is a one-to-one correspondence between Clifford superconnections
and Dirac operators (cf. [BGV]), we present in section 4
formula (3.14) in this context. In the following section, we classify
Dirac operators `of simple type' with respect to the decomposition
of their square. For this kind of Dirac operators, the local
Atiyah-Singer index theorem as proven by Getzler
(cf. [G]) for Dirac operators corresponding to Clifford connections,
also holds (theorem 5.6). In section 6 we turn our attention
to the Wodzicki function $W_{\cal E}$\ defined on the space of all Dirac
operators ${\cal D}({\cal E})$. By the help of our decomposition
formula (3.14) we obtain an explicit expression
for the Wodzicki
function $W_{\cal E}(\widetilde{D})$\ of an arbitrary Dirac operator
$\widetilde{D}$\ in theorem 6.4 . In the last section,
we apply this result to calculate
the Wodzicki function $W_{\cal E}$\
for various Dirac operators. These examples are also
inspired by physics. Note that by this method
we are able to
derive the combined Einstein-Hilbert/Yang-Mills action out of
one (special) Dirac operator. From a physical point of view,
this can be understood as unification of Einstein's gravity and
Yang-Mills gauge theories (cf. [AT2]).
\medskip
For the reader who is less familiar with the notions of Clifford
module, Clifford connection and Dirac operator we recommend the
recent book [BGV] which can also serve as an excellent introduction
into the theory of superconnections and Clifford superconnections.
\vfill\break
{\mittel 2. Dirac operators}
\vskip 0.7cm
\ukneu
Let $M$\ be an even-dimensional Riemannian manifold and ${\cal E}=
{\cal E}^+\oplus {\cal E}^-$\
a $\zz_2$-graded vector bundle over $M$.
A Dirac
operator acting on sections of ${\cal E}$\
is an odd-parity first order operator
$${D\colon \Gamma({\cal E}^\pm)\longrightarrow \Gamma(
{\cal E}^\mp)}\eqno(2.1)$$
such that $D^2$\ is a generalized laplacian. Here we consider
such bundles ${\cal E}$\ provided with a fixed $\zz_2$-graded
left action
${c\colon C(M)\times {\cal E}\rightarrow {\cal E}}$\ of the
Clifford bundle $C(M)$, i.e. Clifford modules. For convenience of
the reader and to fix our conventions we recall that
the Clifford bundle $C(M)$\ is a vector bundle over $M$\ whose
fibre at $x\in M$\ consists of the Clifford algebra $C(T^*_xM)$\
generated by $T^*_xM$\ with respect to the relations
${v\star w +w\star v=-2 g_x(v,w)}$\ for all $v, w\in T^*_xM$.
It is well-known that for an even-dimensional spin manifold
$M$\ any Clifford module ${\cal E}$\ is a twisted bundle
${\cal E}=S\otimes E$. Here $S$\ denotes the spinor bundle
and $E$\ is
a vector bundle with trivial Clifford action uniquely determined
by $\cal E$.
\smallskip
We will regard only those Dirac operators $D$\ that
are compatible with the given Clifford module structure of ${\cal E}$.
This means that
$${[D,f]=c(df)}\eqno(2.2)$$
holds for all $f\in C^\infty(M)$. Property (2.2) fully
characterizes these
Dirac operators : If $P$\ is a differential operator
$P\colon \Gamma({\cal E}^\pm)\rightarrow \Gamma({\cal E}^\mp)$\
with $[P,f]=c(df)$\ for all $f\in C^\infty(M)$, then $P$\ is a
Dirac operator, cf. [BGV].
Given any connection $\nabla^{\cal E}\colon \Gamma({\cal E}^\pm)
\rightarrow \Gamma(T^*M\otimes {\cal E}^\pm)$\ on ${\cal E}$\
which respects the grading, the first-order operator
$D_{\nabla^{\cal E}}$\ defined by
the following composition
$${\Gamma({\cal E}^\pm
)\;\mapright{\nabla^{\cal E}}\; \Gamma(T^*M\otimes
{\cal E}^\pm)\;\hookrightarrow \Gamma(C(M)\otimes {\cal E})\mapright{c}
\Gamma({\cal E}^\mp)}\eqno(2.3)$$
obviously is a Dirac operator. Note that,
in the case of ${\cal E}=S\otimes
E$, the above construction
yields a canonical Dirac operator  with respect to any fixed
connection $\nabla^E$\ on $E$\ by taking
the tensor product connection
$\nabla^{\cal E}:=\nabla^S\otimes \eins_E + \eins_S\otimes \nabla^E$.
Here
$\nabla^S$\ denotes the spin connection on
$S$\ uniquely determined by the metric structure on
$M$.
\smallskip
For our purpose it is important to realize that every Dirac operator
$\widetilde{D}$\ can be constructed as in (2.3). This follows from
\Lemma Let ${\cal E}$\ be a Clifford module over an even-dimensional
manifold $M$\ and $\widetilde{D}\colon \Gamma({\cal E}^\pm)\rightarrow
\Gamma({\cal E}^\mp)$\ be an arbitrary Dirac operator compatible
with the Clifford module structure. Then there
exists a connection $\widetilde{\nabla}^{\cal E}\colon
\Gamma(T^*M\otimes{\cal E}^\pm)\rightarrow
\Gamma({\cal E}^\pm)$\ on ${\cal E}$\ such that
$\widetilde{D}=D_{\widetilde{\nabla}^{\cal E}}:=c\circ
\widetilde{\nabla}^{\cal E}$.
\smallskip
\Proof Let $D_{\nabla^{\cal E}}:=c\circ \nabla^{\cal E}$\ be a Dirac
operator on ${\cal E}$\ constructed as in (2.3). Since
$D_{\nabla^{\cal E}}$\ and $\widetilde{D}$\ are compatible with the
Clifford action on ${\cal E}$\ we obtain
$${[D_{\nabla^{\cal E}}-\widetilde{D}, f]=c(df)-c(df)=0}\eqno(2.4)$$
for all $f\!\in\! C^\infty(M)$. Hence
${D_{\nabla^{\cal E}}-\!\widetilde{D}}$\ can be considered as
a section $A\!\in\! \Gamma(\End^-( {\cal E}))$. Now let the
linear bundle map
$${\nu\colon C(M)\;\longrightarrow T^*M\otimes C(M)}\eqno(2.5)$$
be locally defined by
$\nu\bigl(c(dx^{i_1})\star\cdots\;
\star c(dx^{i_k})\bigr):=dx^{i_1}\otimes c(dx^{i_2})\star\cdots\;
\star c(dx^{i_k})$\ with $i_1< i_2<\cdots\;<i_k<2n$.
Furthermore, we denote by
$\End_{C(M)}{\cal E}$\
the algebra bundle of bundle endomorphisms of ${\cal E}$\
supercommuting with the action of $C(M)$. Then
the composition of this map $\nu$\ with
the canonical isomorphism $\End {\cal E}\cong
C(M)\otimes \End_{C(M)}{\cal E}$\
induces a map
$${\Gamma(\End^-({\cal E}))\;\mapright{\nu}\;\Gamma(T^*M\otimes
\End^+({\cal E}))=\Omega^1(M,\End^+({\cal E}))}\eqno(2.6)$$
which we denote with the same symbol for convinience. Let
$\omega\in\Omega^1(M,\End^+({\cal E}))$\ be the image of
$A\in \Gamma(\End^-( {\cal E}))$\ under this map $\nu$.
Then $\widetilde{\nabla}^{\cal E}:=\nabla^{\cal E}+\omega$\ obviously
defines a new covariant derivative on ${\cal E}$\
which respects the $\zz_2$-grading, and the Dirac operators
$D_{\widetilde{\nabla}^{\cal E}}$\ and $\widetilde{D}$\ coincide.
\QED
\smallskip
We will now turn our attention to a specific class of connections
on a Clifford module $\cal E$, namely the Clifford connections.
Let us recall that a connection $\nabla^{\cal E}\colon \Gamma({\cal
E}^\pm)\rightarrow
\Gamma(T^*M\otimes {\cal E}^\pm)$\ is called a Clifford connection,
if for any $a\in \Gamma(C(M))$\ and $X\in \Gamma(TM)$\ we have
$${[\nabla^{\cal E}_X, c(a)]=c(\nabla_Xa).}\eqno(2.7)$$\nobreak
In this formula, $\nabla$\ denotes the Levi-Civita connection
extended to the Clifford\goodbreak bundle $C(M)$. In the case of
${\cal E}= S\otimes E$, the above mentioned tensor product
connections
$\nabla^S\otimes \eins_E +\eins\otimes \nabla^E$\ are Clifford
connections. Furthermore, applying a partition of unity
argument, one is able to construct a Clifford connection
on every Clifford module $\cal E$. In lemma 2.1
we can therefore chose $\nabla^{\cal E}$\
to be a Clifford connection, i.e.
$${\widetilde{\nabla}^{\cal E}=\nabla^{\cal E} +
\omega.}\eqno(2.8)$$
This is the only property we need in order to prove our generalisation
of the Lichnerowicz formula.
\smallskip
For later use we introduce the following notion:
We call a graduation on the bundle
$\End_{C(M)}({\cal E})$\
with the property
$$\eqalignno{\End^+({\cal E})&\cong \bigl(C(M)^+{\hat\otimes}
\End_{C(M)}^+
({\cal E})\bigr)\;\oplus \bigl(C(M)^-{\hat \otimes}\End_{C(M)}^-({\cal E})
\bigr) & (2.9)\cr
\End^-({\cal E})&\cong \bigl(C(M)^+{\hat\otimes}
\End_{C(M)}^-
({\cal E})\bigr)\;\oplus \bigl(C(M)^-{\hat\otimes}
 \End_{C(M)}^+({\cal E})
\bigr)  &(2.10)\cr}$$
a twisting graduation of ${\cal E}$. Here ${\hat \otimes}$\ denotes
the $\zz_2$-graded tensor product.
In the case of a twistet spin bundle ${\cal E}=S\otimes E$, obviously
any twisting graduation $\End_{C(M)}({\cal E})=
\End_{C(M)}^+({\cal E}) \oplus \End_{C(M)}^-({\cal E})$\
induces a graduation on the twisting bundle $E$\ and conversely.
Since any Clifford module ${\cal E}$\
may be decomposed as $S\otimes E$\ locally, c.f. [BGV], a twisting
graduation
on ${\cal E}$\
corresponds to a graduation of the twisting part $E$.
\vskip 1.0cm
{\mittel 3. The generalized Lichnerowicz formula}
\ukneu
\vskip 0.6cm
If $\nabla^{\cal E}\colon \Gamma({\cal E})\rightarrow \Gamma(T^*M\otimes
{\cal E})$\ is a Clifford connection on the Clifford module
${\cal E}$, there is the well-known decomposition-formula
for the square of the corresponding Dirac operator
${D_{\nabla^{\cal E}}:= c\circ \nabla^{\cal E}}$\ due to Lichnerowicz [L]:
$${D_{\nabla^{\cal E}}^2=\triangle^{\nabla^{\cal E}}+{r_M\over 4} +
{\bf c}(R^{{\cal E}/S}_{\nabla^{\cal E}}).}\eqno(3.1)$$
Here $\triangle^{\nabla^{\cal E}}$\ denotes the connection
laplacian associated to $\nabla^{\cal E}$\fussnote{${^{(1)}}$}{
With respect
to a local coordinate frame of $TM$, the connection laplacian
${\triangle^{\nabla^{\cal E}}}$\
is explicitly
given by ${\triangle^{\nabla^{\cal E}}=-g^{\mu\nu}(\nabla^{\cal E}_\mu
\nabla^{\cal E}_\nu -\Gamma^\sigma_{\mu\nu}\nabla^{\cal E}_\sigma)}$.},
$r_M$\ is the
scalar curvature of $M$\ and ${{\bf c}(R^{{\cal E}/S}_{
\nabla^{\cal E}})}$\ denotes
the image of the twisting curvature $R^{{\cal E}/S}_{\nabla^{\cal E}}
\in \Omega^2(M,
\End_{C(M)}({\cal E}))$\ of the Clifford connection $\nabla^{\cal E}$\
with respect to
the quantisation map
${{\bf c}\colon \Lambda^*T^*M\rightarrow C(M)}$, cf. [BGV].
\smallskip
In this section we will generalize formula (3.1) for an arbitrary
Dirac operator $\widetilde{D}$\ on ${\cal E}$\ which is
compatible with the Clifford action. Using $\widetilde{D}=c(dx^\mu)
\widetilde{\nabla}^{\cal E}_\mu$\ and mimicking
the first step in the computation of the Lichnerowicz formula (3.1),
we obtain
$${\widetilde{D}^2= -g^{\mu\nu}\widetilde{\nabla}^{\cal E}_\mu\widetilde{
\nabla}^{\cal E}_\nu
+ c(dx^\mu)[\widetilde{\nabla}^{\cal E}_\mu,c(dx^\nu)]
\widetilde{\nabla}^{\cal E}_\nu
+ {1\over 2}\;c(dx^\mu)c(dx^\nu) [\widetilde{\nabla}^{\cal E}_\mu,
\widetilde{\nabla}^{\cal E}_\nu].}\eqno(3.2)$$
It is remarkable that none of the first two terms
in (3.2) is globally defined, but only their
sum. However, we observe that given a Clifford connection
$\nabla^{\cal E}$\ on ${\cal E}$, then
(3.2) can be rearranged in such a way
that ${\widetilde{D}^2=\triangle^{\nabla^{\cal E}}+ P+ F^\prime}$\
where $P\colon \Gamma({\cal E})\rightarrow \Gamma({\cal E})$\ is
a first-order differential operator and $F^\prime \in \End({\cal E})$\
with both terms
depending
on $(\widetilde{\nabla}^{\cal E}-\nabla^{\cal E})=:\omega\in
\Omega^1(M,\End^+({\cal E}))$. More precisely we have
\Lemma Let ${\nabla^{\cal E}\colon \Gamma(
{\cal E}^\pm)\rightarrow \Gamma(T^*M\otimes {\cal E}^\pm)}$\ be a
Clifford connection on the Clifford module $\cal E$\ and
$\widetilde{D}=c\circ \widetilde{\nabla}$\ a Dirac operator.
If $\omega$\ denotes the one-form on $M$\ with values in $\End^+({\cal E})$\
uniquely determined by $\omega:=\widetilde{\nabla}-\nabla^{\cal E}$, then
$${\widetilde{D}^2=\triangle^{\nabla^{\cal E}} - (B_\omega\nabla^{\cal E})
+{r_M\over 4} +
{\bf c}(R^{{\cal E}/S}_{\nabla^{\cal E}}) +F_\omega^\prime}\eqno(3.3)$$
where $B_\omega\colon \Gamma(T^*M\otimes {\cal E})\rightarrow
\Gamma({\cal E})$\
with $B_\omega(dx^\nu\otimes s) =\bigl(2g^{\mu\nu}\omega_\mu -c(dx^\mu)[
\omega_\mu, c(dx^\nu)]\bigr)s$\
and ${F_\omega^\prime\!=\!c(dx^\mu)c(dx^\nu)\bigl(\;^\prime\nabla_\mu
\omega_\nu\bigr) \!+\!
c(dx^\mu)\omega_\mu c(dx^\nu)
\omega_\nu}$\ together with the Lichnerowicz terms
${{r_M\over 4} +
{\bf c}(R^{{\cal E}/S}_{\nabla^{\cal E}})}$\
determine the first order resp.
the endomorphism part with respect to a local coordinate system.
\smallskip
\Proof By inserting ${\widetilde{\nabla}=\nabla^{\cal E} +
\omega}$\
in (3.2) and by using that
$\nabla^{\cal E}$\ is a Clifford connection, we get
$$\eqalign{\widetilde{D}^2 &=\triangle^{\nabla^{\cal E}}
 +{ 1\over 4}\; [c(dx^\mu),c(dx^\nu)]\;[\nabla^{\cal E}_\mu,\nabla^{\cal
E}_\nu]-
g^{\mu\nu}\bigl([\nabla^{\cal E}_\mu,\omega_\nu]
-\Gamma^\sigma_{\nu\mu}\omega_\sigma\bigr)\cr
&\ \  -g^{\mu\nu}\omega_\mu \omega_\nu
 -g^{\mu\nu}\bigl(2 \omega_\mu\nabla^{\cal E}_\nu\bigr)
 + c(dx^\mu)[\omega_\mu,c(dx^\nu)]
\nabla^{\cal E}_\nu \cr
&\ \ +{{\petit\rm 1}\over {\petit 4}}\;[c(dx^\mu),c(dx^\nu)
]\;
[\omega_\mu,\omega_\nu]
+{1\over 2}[c(dx^\mu), c(dx^\nu)][\nabla^{\cal E}_\mu,\omega_\nu] \cr
&\ \ +c(dx^\mu)[\omega_\mu,c(dx^\nu)]
\omega_\nu\cr}\eqno(3.4)$$
with respect to the local coordinate frame
$\{\partial_\mu\}$\ of $TM$.,
Here $\Gamma^\sigma_{\mu\nu}$\ are the Christoffel symbols
defined by the Levi-Civita connection on $M$. Thus,
the fifth together with the sixth term on the
right-hand-side define $B_\omega\colon \Gamma(T^*M\otimes {\cal E})\rightarrow
\Gamma({\cal E})$. Furthermore we have to explain our `short-hand'
notation
${(^\prime\nabla_\mu\omega_\nu)}$:
$$\eqalign{-g^{\mu\nu}\bigl([\nabla^{\cal E}_\mu,\omega_\nu]-
\Gamma^\sigma_{\mu\nu}\omega_\sigma\bigr)
&= -g^{\mu\nu}\bigl(\nabla^{{\petit\rm End}\;{\cal E}}_\mu \omega_\nu)-
\Gamma^\sigma_{\mu\nu}\omega_\sigma\bigr) =:
-g^{\mu\nu}(^\prime\nabla_\mu \omega_\nu)\cr
{1\over 2} [c(dx^\mu),c(dx^\nu)][\nabla^{\cal E}_\mu,
\omega_\nu] &={1\over 2} [c(dx^\mu),c(dx^\nu)]\bigl(
\nabla^{{\petit\rm End}\;{\cal E}}_\mu \omega_\nu) -\Gamma^\sigma_{\mu\nu}
\omega_\sigma\bigr)\cr
 &=:
{1\over 2} [c(dx^\mu),c(dx^\nu)](^\prime\nabla_\mu \omega_\nu).\cr}\eqno(3.5)$$
Equivalently one could describe (3.5) in a global manner by
using the composition of the connection
${\nabla^{T\!^*\!M\otimes {\petit\rm End}{\cal E}}\colon
\Gamma(T^*M\!\otimes\!\End {\cal E})\rightarrow
\Gamma(T^*M\!\otimes\! T^*M\!\otimes\!\End {\cal E})}$\
together with the evaluation $-ev_g$\ and the quantisation map
{\bf c}, respectively. Finally we use ${1\over
2}[c(dx^\mu),c(dx^\nu)]-g^{\mu\nu}
=c(dx^\mu)c(dx^\nu)$\ to obtain the first term of $F^\prime\in \Gamma(
\End({\cal E}))$. The computation
of the second term of $F^\prime$\ is straightforward.
\QED
\smallskip
Of course, in spite of being global, the decomposition formula
(3.3) seems to be unsatisfactory since the first-order
operator $B_\omega\nabla^{\cal E}$\ does not vanish,
generally\fussnote{${^{(2)}}$}{
We will study the case $B_\omega=0$\ in the next section.}.
Also (3.3) term by term depends
on the chosen Clifford connection $\nabla^{\cal E}$\ and is therefore
by no means an intrinsic property of the Dirac operator $\widetilde{D}$.
This shows that (3.3)
can not serve as a generalisation of the
Lichnerowicz formula (3.1).
To remedy this flaw we need the following observation:
\Lemma Let ${\triangle^{\nabla^E}}$\ and ${\triangle^{{\hat\nabla}^E}}$\
be the connection laplacians defined with respect to the connections
${\nabla^E}$\ and ${{\hat\nabla}^E}$\ acting on sections of
a vector bundle $E$\ over $M$. If furthermore  $({\hat\nabla}^E-\nabla^E)
=:\alpha\in \Omega^1(M,\End\;E)$, then
$${\triangle^{{\hat\nabla}^E}=\triangle^{\nabla^E}-
A_\alpha\nabla^E-\Theta_\alpha,}\eqno(3.6)$$
where $A_\alpha\colon \Gamma(T^*M\otimes E)\rightarrow \Gamma(E)$\
with ${A_\alpha(dx^\nu\otimes s_\nu):=2 g^{\mu\nu}\alpha_\mu s_\nu}$\
and the endomorphism $\Theta_\alpha:=g^{\mu\nu}(\;^\prime\nabla^E_\mu
\alpha_\nu)+ g^{\mu\nu}\alpha_\mu\alpha_\nu$\ are defined with respect
to a local coordinate system.
\smallskip
\Proof Obviously we have ${\hat \nabla}^E=\nabla^E+\alpha$\ and
compute
$$\eqalign{\triangle^{{\hat\nabla}^E} :&= - g^{\mu\nu}\bigl(
(\nabla^E_\mu+\alpha_\mu)(\nabla^E_\nu+\alpha_\nu) -
\Gamma^\sigma_{\mu\nu}(\nabla^E_\sigma +\alpha_\sigma)\bigr)\cr
&=\underbrace{-g^{\mu\nu}(\nabla^E_\mu\nabla^E_\nu -\Gamma^\sigma_{\mu\nu}
\nabla^E_\sigma)}_{=\;\triangle^{\nabla^E}}\cr
&\ \ -\underbrace{
2g^{\mu\nu}\alpha_\mu\nabla_\nu}_{=\;A_\alpha\nabla^E} - \underbrace{
g^{\mu\nu}\bigl(
(\nabla^{{\petit \rm End}\;E}_\mu\alpha_\nu-\Gamma^\sigma_{\mu\nu}
\alpha_\sigma)+\alpha_\mu\alpha_\nu\bigr)}_{=\;\Theta_\alpha}\;,\cr
}\eqno(3.7)$$
where we have also used the definition of ${^\prime\nabla^E}$\
given in (3.5).
\QED
\smallskip
Note, that $A_\alpha\in \Gamma(TM\otimes\End\;E)$\ can be described also
by $A_\alpha =2\;\alpha^\sharp$, where $\alpha^\sharp$\ denotes the
corresponding `dual' element of $\alpha\in \Gamma(T^*M\otimes\End\;E)$\
in $\Gamma(TM\otimes\End\;E)$\ under the `musical' isomorphism
${T^*M\otimes \End\;E \;\displaystyle{\hir^{\sharp\otimes \eins}_{
\flat\otimes \eins}}\;
TM\otimes \End\;E}$\ defined by the Riemannian metric $g$.
\smallskip
Guided by the previous lemma in the case of $E={\cal E}$\ and
$\nabla^E=\nabla^{\cal E}$\ the given Clifford connection,
we define
${{\hat\nabla}^{\cal E}:=\nabla^{\cal E}+ {1\over 2}\;
B_\omega^\flat}$,
with $B_\omega\in \Gamma(TM\otimes\End {\cal E})$\ as above.
Thus we have
${{\hat\nabla}^{\cal E}=\widetilde{\nabla}^{\cal E}
+{\varpi}}$\ with
$${\varpi
:=-{1\over 2}({\partial\over \partial x^\nu}\otimes c(dx^\mu)
[\omega_\mu,
c(dx^\nu)])^\flat=-{1\over 2} g_{\sigma\nu}dx^\sigma\otimes c(dx^\mu)
[\omega_\mu,c(dx^\nu)].}\eqno(3.8)$$
In fact, this one form $\varpi$\ `measures' how much
$\widetilde{\nabla}^{\cal E}$\ differs from beeing a Clifford
connection because ${\varpi=
-{1\over 2} g_{\sigma\nu}dx^\sigma\otimes
c(dx^\mu)\bigl([(\widetilde{\nabla}^{\cal E}_\mu,
c(dx^\nu)]-[\nabla^{\cal E}_\mu,
c(dx^\nu)]\bigr)}$. Since
any two Clifford connections on the Clifford module
${\cal E}$\
differ by an element of $\Omega^1(M,\End_{C(M)}({\cal E}))$,
$\varpi$\ only depends on the initial connection
$\widetilde{\nabla}^{\cal E}$\fussnote{${^{(3)}}$}{In the further
we will therefore
write $\varpi_{\widetilde{\nabla}^{\cal E}}$\ to indicate this
dependence.}. Consequently also ${\hat\nabla}^{\cal E}$\
only depends on $\widetilde{\nabla}^{\cal E}$.
So we get as a result of the previous lemmas:
\Lemma Let ${\cal E}$\ be a Clifford module over an even
dimensional Riemannian manifold $M$\ and
$\widetilde{D}=c\circ \widetilde{\nabla}^{\cal E}$\ a Dirac operator.
If $\triangle^{{\hat\nabla}^{\cal E}}$\ denotes the connection
laplacian corresponding to the above defined connection
${\hat\nabla}$, then
${\widetilde{D}^2\!=\!\triangle^{{\hat\nabla}^{\cal E}}
\! +\! F_{\widetilde{\nabla}^{\cal E}}}$,\nobreak
where
$F_{\widetilde{\nabla}^{\cal E}}\in \End\;({\cal E})$\ is given
by ${F_{\widetilde{\nabla}^{\cal E}}={r_M\over 4}+
{\bf c}(R^{{\cal E}/S}_{\nabla^{\cal E}})
+F^\prime_\omega+\Theta_{{1\over 2}\!
B_\omega^\flat}}$\ with respect to an arbitrary Clifford connection
$\nabla^{\cal E}$\ and $\omega:=(\widetilde{\nabla}^{\cal E}-
\nabla^{\cal E})$.\goodbreak
\Proof By lemma 3.1 we can decompose the square of the Dirac operator
$\widetilde{D}$\ as
$${\widetilde{D}^2=\triangle^{\nabla^{\cal E}} - (B_\omega\nabla^{\cal E})
+{r_M\over 4} +
{\bf c}(R^{{\cal E}/S}_{\nabla^{\cal E}}) +F_\omega^\prime .}\eqno(3.9)$$
When applying lemma 3.2 to the two connection laplacians
$\triangle^{\nabla^{\cal E}}$\ and
${\triangle^{{\hat
\nabla}^{\cal E}}}$,
which correspond
to the given Clifford connection ${\nabla^{\cal E}}$\ and the
above defined connection ${\hat\nabla}:= \nabla^{\cal E}+
{1\over 2} B^\flat_\omega$\
with $\omega:=(\widetilde{\nabla}^{\cal E}
- \nabla^{\cal E})$\ as in lemma 3.1, we obtain
$${\triangle^{\nabla^{\cal E}}-B_\omega\nabla^{\cal E}=
\triangle^{{\hat\nabla}^{\cal E}}+\Theta_{{1\over 2}\!
B^\flat_\omega},}\eqno(3.10)$$
which implies the desired result.
\QED
\smallskip
Since the Dirac operator $\widetilde{D}$\ as well as
the connection laplacian $\triangle^{{\hat\nabla}^{\cal E}}$\ only depend
on $\widetilde{\nabla}^{\cal E}$, obviously so does the
endomorphism $F_{\widetilde{\nabla}^{\cal E}}$. This suggests an
intrinsic meaning of this term. Using the definitions
$$\eqalign{F_\omega^\prime &:=c(dx^\mu)c(dx^\nu)(^\prime\!
\nabla^{\cal E}_\mu
\omega_\nu)+c(dx^\mu)\omega_\mu c(dx^\nu)\omega_\nu\cr
{1\over 2} B^\flat_\omega &:=\omega -
{1\over 2} g_{\sigma\nu}dx^\sigma\otimes c(dx^\mu)
[\omega_\mu,c(dx^\nu)] \cr
\Theta_{{1\over 2}\!B^\flat_\omega} &:= {1\over 2}\;g^{\mu\nu}
\bigl(^\prime\!\nabla^{\cal E}_\mu (B^\flat_\omega)_\nu\bigr) +
{1\over 4}\;g^{\mu\nu}(B^\flat_\omega)_\mu
(B^\flat_\omega)_\nu\cr}\eqno(3.11)$$
together with the identity ${{r_M\over 4}+ c(R^{{\cal E}/S}_{
\nabla^{\cal
E}})={1\over 4}[c(dx^\mu) , c(dx^\nu)][\nabla^{\cal E}_\mu,
\nabla^{\cal E}_\nu]}$\ we compute
$$\eqalign{F_{\widetilde{\nabla}^{\cal E}}=&
{{\petit\rm 1}\over {\petit\rm 4}}\;
[c(dx^\mu),c(dx^\nu)][
\nabla^{\cal E}_\mu ,\nabla^{\cal E}_\nu]
+{1\over 2}[c(dx^\mu),c(dx^\nu)]\bigl(^\prime\!\nabla^{\cal E}_\mu
\omega_\nu\bigr)\cr
&\ \ + {1\over 4}[c(dx^\mu),c(dx^\nu)][\omega_\mu,\omega_\nu]
+ g^{\sigma\nu}[\widetilde{\nabla}^{\cal E}_\nu,\;-{1\over 2}
g_{\sigma\nu}c(dx^\mu)
[\omega_\mu,c(dx^\nu)]] \cr
&\ \ +{1\over 4} g_{\mu\nu}c(dx^\kappa)[
\omega_\kappa, c(dx^\mu)]c(dx^\sigma)[\omega_\sigma, c(dx^\nu)].
\cr}\eqno(3.12)$$
The first three terms of (3.12) join together to
define ${1\over 4}[c(dx^\mu),c(dx^\nu)]
[\widetilde{\nabla}^{\cal E},\widetilde{\nabla}^{\cal E}]=
{\bf c}(R^{\widetilde{\nabla}^{\cal E}})$, i.e. the
image of the curvature $R^{\widetilde{\nabla}^{\cal E}}\in
\Omega^2(M,\End\;{\cal E})$\ of the connection $\widetilde{\nabla}^{\cal
E}$\ under the quantization map {\bf c}. The forth term can be
written as
${ev_g\widetilde{\nabla}^{T^*M\otimes {\petit\rm End} {\cal E}}
\varpi_{\widetilde{
\nabla}^{\cal E}}}$\
with the tensor product connection
${\widetilde{\nabla}^{T^*M\otimes {\petit\rm End}{\cal E}}:=
\nabla\otimes \eins_{\cal E} + \eins_{T^*M}\otimes
\widetilde{\nabla}^{\cal E}}$, cf. equation (3.5) and the
remark thereafter. Using the pointwise defined product `$\cdot$'
in the algebra bundle
$T(M)\otimes \End {\cal E}$, where $T(M)$\ denotes the tensor
bundle of $T^*M$, we can write
$ev_g(\varpi_{\widetilde{\nabla}^{\cal E}}
\cdot \varpi_{\widetilde{\nabla}^{\cal E}}
)$\ for the last term. So we get
$${F_{\widetilde{\nabla}^{\cal E}}=
{\bf c}(R^{\widetilde{\nabla}^{\cal
E}})+
ev_g\widetilde{\nabla}^{T^*M\otimes {\petit\rm End}{\cal E}}
\varpi_{\widetilde{\nabla}^{\cal E}} +
ev_g(\varpi_{\widetilde{\nabla}^{\cal E}}
\cdot \varpi_{\widetilde{\nabla}^{\cal E}}).}\eqno(3.13)$$
Therefore we have shown
\Theorem Let ${\cal E}$\ be a Clifford module over an even
dimensional Riemannian manifold $M$\ and
$\widetilde{D}=c\circ \widetilde{\nabla}^{\cal E}$\ a Dirac operator.
Then
$${\widetilde{D}^2=\triangle^{{\hat\nabla}^{\cal E}}+
{\bf c}(R^{\widetilde{\nabla}^{\cal
E}})+
ev_g\widetilde{\nabla}^{T^*M\otimes {\petit\rm End}{\cal E}}
\varpi_{\widetilde{\nabla}^{\cal E}} +
ev_g(\varpi_{\widetilde{\nabla}^{\cal E}}
\cdot \varpi_{\widetilde{\nabla}^{\cal E}}),}\eqno(3.14)$$
with $\varpi_{\widetilde{\nabla}^{\cal E}}:=-{1\over 2} g_{\nu\kappa}
dx^\nu\otimes
c(dx^\mu)\bigl([\widetilde{\nabla}^{\cal E}_\mu, c(dx^\kappa)] +
c(dx^\sigma)\Gamma^\kappa_{\sigma\mu}\bigr)\in \Omega^1(M,\End {\cal E})$\
and the connection ${\hat\nabla}^{\cal E}:=\widetilde{\nabla}^{\cal E}+
\varpi_{\widetilde{\nabla}^{\cal E}}$.
\smallskip\rm
In this decomposition of the square of $\widetilde{D}$, the last two
terms obviously indicate the deviation of the
connection $\widetilde{\nabla}^{\cal
E}$\ being a Clifford connection. Only the second term
of (3.14) is endowed with geometric significance. Of course, if
$\widetilde{\nabla}^{\cal E}$\
is a Clifford connection,
obviously $\varpi_{\widetilde{\nabla}^{\cal E}}=0$\ and therefore
(3.14) reduces to Lichnerowicz's formula
${\widetilde{D}^2\!=\!\triangle^{\widetilde{\nabla}^{\cal E}}\!+\!
{r_M\over 4}\! +\!
{\bf c}(\!R^{{\cal E}/S}_{\!\widetilde{\nabla}^{\cal E}}\!)}$. So we call
(3.14) `the generalized Lichnerowicz formula'.
\vskip 1.0cm
{\mittel 4. The `Super - Lichnerowicz formula'}
\vskip 0.6cm
\ukneu
In this section we will turn our attention to the generalized
Lichnerowicz formula (3.14) in the context of `super-geometry'.
This is motivated by
the well-known fact, that any Clifford superconnection
$A$\ on a Clifford module $\cal E$\ uniquely determines a
Dirac operator ${D_A}$\
due to the following construction
$${D_A\colon \Gamma({\cal E})\;\mapright{A}\;\Omega^*(M,{\cal E})\;
\mapright{{\petit\bf c}\otimes \eins_{\cal E}}\;\Gamma(C(M)\otimes
{\cal E})\;\mapright{c}\; \Gamma({\cal E}), }\eqno(4.1)$$
i.e. there is a one-to-one correspondence between Clifford
superconnections and
Dirac operators, see [BGV]\fussnote{${^{(4)}}$}{This reference
can serve also as an excelent introduction into the
theory of superconnections and Clifford superconnections.}.
Since (3.14) is already the decomposition of
the square of an arbitrary Dirac operator on $\cal E$,
we only have to adapt it to
super-geometry. Thus, the expression `Super-Lichnerowicz formula' is
an abuse of notation here. To proceed,
let again $\nu\colon C(M)\rightarrow T^*M\otimes C(M)$\ be the
linear bundle
map defined in the second section. The following simple observation
is crucial:
\Lemma Let $A$\ be a Clifford superconnection on the Clifford
module ${\cal E}$. Then the operator $\nabla^A$\ defined by the
following composition
$${\Gamma({\cal E})\mapright{A} \Omega^*(M, {\cal E})\mapright{
{\petit\bf c}\otimes \eins_{\cal E}} \Gamma(C(M)\otimes
{\cal E})\mapright{\nu\otimes \eins_{\cal E}}
\Gamma(T^*M\otimes C(M)\otimes {\cal E})
\mapright{\eins_{T^*M}\otimes c}\Gamma(T^*M\otimes {\cal E})}$$
is a connection on ${\cal E}$\
with the property that the associated Dirac operator $D_{\nabla^A}:=
c\circ \nabla^A$\ coincides with $D_A$.
\smallskip
\Proof Because $A$\ satisfies Leibniz's rule
$A(fs)=
df\otimes s + fA s$\ for all $f\in C^\infty({\cal E})$\ and
$s\in \Gamma({\cal E})$, so does $\nabla^A$. Therefore
$\nabla^A$\ is a connection on ${\cal E}$.
To prove the second statement we simply remark, that the
diagramm
$${\matrix{&\Gamma(C(M)\otimes {\cal E}) &\mapright{c} &\Gamma({\cal E})
\cr
\noalign{\vskip 0.1cm}
&\mapdown{\nu\otimes \eins_{\cal E}} &\quad &\mapup{c}\cr
\noalign{\vskip 0.1cm}
&\Gamma(T^*M\otimes C(M)\otimes {\cal E}) &\mapright{\eins_{T^*M}
\otimes c} &\Gamma(T^*M\otimes {\cal E}) \cr} }\eqno(4.2)$$
commutes.\nobreak\QED\goodbreak
\smallskip
In general, however, the connection $\nabla^A$\ defined by a Clifford
superconnection is not a Clifford connection, i.e.
${[\nabla^A_\mu, c(dx^\nu)]\ne -c(dx^\sigma)\Gamma^\nu_{\sigma\mu}}$.
Using lemma 4.1 , we can now `reformulate' theorem 3.4 in the following
way
\Theorem Let ${\cal E}$\ be a Clifford module over an even
dimensional Riemannian manifold $M$, $A\colon\Omega^\pm(M,{\cal E})
\rightarrow \Omega^\mp(M,{\cal E})$\ a Clifford superconnection
and $D_A$\ the corresponding Dirac operator. Then
$${D_A^2=\triangle^{{\hat\nabla}^{\cal E}}+
{\bf c}(R^{\nabla^A})+
ev_g({\nabla^A})^{T^*M\otimes {\petit\rm End}{\cal E}}
\varpi_{\nabla^A} +
ev_g(\varpi_{\nabla^A}
\cdot \varpi_{\nabla^A}),}\eqno(4.3)$$\nobreak
with $\varpi_{\nabla^A}:=-{1\over 2} g_{\nu\kappa}
dx^\nu\otimes
c(dx^\mu)\bigl([\nabla^A_\mu, c(dx^\kappa)] +
c(dx^\sigma)\Gamma^\kappa_{\sigma\mu}\bigr)\in \Omega^1(M,
\End {\cal E})$\
and the connection ${\hat\nabla}^{\cal E}:=\nabla^A +
\varpi_{\nabla^A}$.\goodbreak
{\mittel 5. Dirac operators of simple type}
\rm
\ukneu
\vskip 0.6cm
Now we will begin with our analysis of Dirac operators
with regard to our
decomposition formula (3.14). We start with the following
\Definition A Dirac operator $\widetilde{D}$
acting on sections of a Clifford module
${\cal E}$\ is called `of simple type'
if the connection ${{\hat\nabla}^{\cal E}}$\
which defines the connection laplacian
${\triangle^{{\hat\nabla}^{\cal E}}}$\ in the decomposition formula
(3.14) of $\widetilde{D}^2$\
is a Clifford connection.
\smallskip\rm
Because it is essentially the section $B_\omega\in \Gamma(TM
\otimes \End {\cal E})$\ which defines the connection
${\hat\nabla}^{\cal E}$\ (cf. the definition of
${\hat\nabla}^{\cal E}$\ before (3.8)), such Dirac operators
are
characterized by the following
\Lemma Let $\widetilde{D}$\ be a Dirac operator on a Clifford module
${\cal E}$. Then $\widetilde{D}$\ is of simple type iff
there exists a Clifford connection ${\nabla^{\cal E}\colon
\Gamma({\cal E})\rightarrow \Gamma(T^*M\otimes {\cal E})}$\
such that $B_\omega\in \Gamma(TM\otimes \End {\cal E})$\
in the decomposition (3.3) of $\widetilde{D}^2$\ vanishes.
\smallskip\rm
Obviously any Dirac operator $D_{\nabla^{\cal E}}$\
associated with a Clifford connection $\nabla^{\cal E}$\
is of simple type. Now we ask wether there are still
other Dirac operators of simple type on ${\cal E}$. If the
twisting graduation of the Clifford module ${\cal E}$\ is non-trivial,
we will answer
this question affirmatively. Moreover we will show, that Dirac operators
of simple type are in one-to-one correspondence with
the set of pairs $\{(\nabla^{\cal E},\phi)\}$. Here $\nabla^{\cal E}$\
denotes a Clifford connection on the respective Clifford module
${\cal E}$\ and $\phi$\ is a
section of the endomorphism bundle ${\End_{C(M)}^-({\cal E})}$.
In the following we denote by
$Sym^2(T^*M)$\ the bundle of
symmetric two-tensors of $T^*M$\ over $M$. We further remark that
there is a natural inclusion $Sym^2(T^*M)\hookrightarrow T^*M\otimes
C^-(M)$.
\smallskip
\Lemma Let ${\cal E}$\ be a Clifford module over an even dimensional
Riemannian manifold $M$\ and let the linear map
${\it \Xi}\colon
\Omega^1(M,\End^+({\cal E}))\rightarrow \Gamma(TM\otimes \End(
{\cal E}))$\
be defined by $\omega\mapsto B_\omega$.
Then $\ker \;{\it \Xi}\ne 0$\ iff the twisting graduation
of ${\cal E}$\ is non-trivial. Moreover $\omega\in \ker\;
{\it\Xi}$\ iff $\omega=w\otimes
F$\ with $w\in \Gamma(Sym^2(T^*M))$\ and $F\in \Gamma(\End_{C(M)}^-(
{\cal E}))$.
\smallskip
\Proof On an open subset
$U\subset M$\ any $\omega\in \Omega^1(M,\End^+({\cal E}))$\ can be
described by
$${\omega={\sum_{\vert I\vert}}^\prime\;\omega^i_{\mu\;I} dx^\mu
\otimes c(dx^I){\hat\otimes} F_i,}\eqno(5.1)$$
where $\omega^i_{\mu\;I}\in C^\infty(U)$\ and $F_i\in \End_{C(M)}({\cal
E})$, $I=(i_1,\cdots \; i_{\vert I\vert})$\ is a multi-index, $\vert\;
I\;\vert$\
denotes its length, $c(dx^I):=c(dx^{i_1})\cdots\;c(dx^{i_l})$\
and $\sum^\prime$\ signifies that the sum is taken over
strictly increasing indices\fussnote{${^{(4)}}$}{Without
using the Einstein summation convention this means
$${{\sum_{\vert I\vert}}^\prime
\;\omega^i_{\mu\;I} dx^\mu
\otimes c(dx^I){\hat\otimes} F_i:=
\sum_{\vert I\vert}\;
\sum_{\mu,\ i \atop{i_1<\dots <i_{\vert I\vert}}}
\;\omega^i_{\mu\;I} dx^\mu
\otimes c(dx^I){\hat\otimes} F_i.}$$}. By the definition
of $B_\omega$\ (see lemma 3.1) we get
$$\eqalign{B_\omega^\nu &s_\nu= B(dx^\nu \otimes s_\nu)=
{\sum_{\vert I\vert}}^\prime\bigl(2g^{\mu\nu}\!\omega^i_{\mu I}\!
c(\!dx^I\!)\!{\hat \otimes}F_i s_\nu\cr
&\ -{\sum_{\vert I\vert}}^\prime
\omega^i_{\mu I}\!((-1)^{\!\vert\! F_i\!\vert \vert\!
c(\!dx^\nu\!)
\!\vert}\!
c(\!dx^\mu\!)c(\!dx^I\!)c(\!dx^\nu\!)\!-\! c(\!dx^\mu\!)c(\!dx^\nu\!)
c(\!dx^I\!)\bigr)\!{\hat\otimes}\!
F_i s_\nu.\cr}\eqno(5.2)$$
Here the sign $(-1)^{\vert F_i\vert\vert c(dx^\nu)\vert}$\
is a consequence of using the $\zz_2$-graded tensor product in
(5.1). We have to determine the solutions of the
equation $B_\omega^\nu=0$\ for all
$\nu\in \{1,\cdots\; 2n\}$.
\medskip
$\underline{1.\ \rm case}$: $\End^+({\cal E})\cong C^+(M)\otimes
\End_{C(M)}({\cal E})\cong C^+(M){\hat\otimes}
\End_{C(M)}^+({\cal E})$, i.e. the twisting graduation of
${\cal E}$\ is trivial.
\smallskip
Using $F_i\in \End_{C(M)}^+({\cal E})$, i.e.
$\vert F_i\vert=0$\ for all $i$,
together with the Clifford relation ${c(dx^{i_k})c(dx^\nu)+
c(dx^\nu)c(dx^{i_k}))=-2g^{i_k \nu}}$\ we obtain from (3.8)
$${{\sum_{\vert I\vert\;\petit\rm even}}\!^\prime\; \Bigl(
g^{\mu\nu}\omega^i_{\mu I}
c(dx^I) +(-1)^k g^{i_k\nu}\omega^i_{\mu I}
c(dx^\mu)c(dx^{i_1})\dots \widehat{ c(dx^{i_k})}\dots c(dx^{i_{\vert I\vert}}
)\Bigr) =0,}$$
where the hat indicates that the corresponding factor has been
omitted. In the following we supress the endomorphism-index
$i$\ for convinience.
Since each coefficient
$\omega_{\mu I}=\omega_{\mu i_1\dots\;
i_{\vert I\vert}}$\ is totally antisymmetric in the multi-index $I$,
we have $(-1)^k \omega_{\mu i_1\dots i_k\dots\; i_{\vert I\vert
}}=-\omega_{\mu i_k i_1\dots\widehat{i_k}\dots\; i_{\vert
I\vert}}$. With the definition of $\sum^\prime_{\vert I\vert}$\
and using the short-hand notation
$c(dx^{i_1\dots\widehat{i_k}\dots i_{\vert I\vert}}):=
c(dx^{i_1})\dots \widehat{ c(dx^{i_k})}\dots c(dx^{i_{\vert
I\vert}})$, we therefore obtain for the second term above:
$$\eqalign{\!\!{\sum_{\vert I\vert\; \petit\rm even}}\hskip -0.15cm ^\prime
(-1)^k\! g^{i_k\!\nu}\!\omega_{\mu I}\!
c(\!dx^\mu\!)\!c(\!dx^{\!i_1\dots \widehat{i_k}\dots i_{\vert I\vert}}\!)
&=\!\!\hskip -0.3cm\sum_{{\mu \atop i_1<\dots <i_{\vert I\vert} } \atop
{\vert I\vert\; \petit\rm even} }\!\!\hskip -0.3cm
(-1)^k\! g^{i_k\!\nu}\!\omega_{\mu i_1\dots i_k\dots
i_{\vert I\vert}}\!
c(\!dx^\mu\!)c(\!dx^{i_1\dots \widehat{i_k}\dots i_{\vert I\vert}}\!)
\cr
&=\!\!\sum_{{\mu \atop i_2<\dots <i_{\vert I\vert} }\atop{
\vert I\vert \;\petit\rm even}}\!\hskip -0.2cm -
\Bigl({1\over \vert I\vert}\;
{{\omega_\mu}^\nu}_{i_2\dots i_{\vert I\vert}}c(dx^\mu)c(dx^{i_2\dots
i_{\vert I\vert}})\Bigr)\cr
&=\!\!\sum_{\mu<i_2<\dots <i_{\vert I\vert}\atop \vert I\vert\;
\petit\rm even}\!\! \hskip -0.2cm -
{{\omega_\mu}^\nu}_{i_2\dots i_{\vert I\vert}}c(dx^{\mu i_2\dots
i_{\vert I\vert}})\cr
&\ \ \ \ + \sum_{{\mu \atop i_3<\dots i_{\vert I\vert}}\atop
{\vert
I\vert\;\petit\rm even}}\!\!
{\omega_\mu}^\nu_{\phantom{\nu} \mu i_3\dots i_{\vert I\vert}}
c(dx^{i_3\dots i_{\vert I\vert}}).\cr}$$
Consequently for
$\mu=i_1$\ we have to solve
$${{\sum_{\vert I\vert\;\petit\rm even}}\hskip -0.15cm ^\prime\;\bigl(
{\omega^\nu}_{i_1\dots i_{\vert I\vert}} -
{{\omega_{i_1}}^\nu}_{i_2\dots i_{\vert I\vert}}\bigr) c(dx^I)
+ \sum_{{i_1 \atop i_3<\dots i_{\vert I\vert}}\atop
{\vert
I\vert\;\petit\rm even}}\!\!
{\omega_{i_1}}^\nu_{\phantom{\nu} i_1  i_3\dots i_{\vert I\vert}}
c(dx^{i_3\dots i_{\vert I\vert}}) =0}\eqno(5.3)$$
for all $1\le\nu\le 2n$.
Now let $\vert I\vert>0$. Since the indices
are ordered and therefore the vectors $c(dx^I)$\ resp.
$c(dx^{i_3\dots i_{\vert I\vert}})$\ are linear independent in
the Clifford algebra, equation
(5.3) shows that
$${\omega_{i_1\nu i_2\dots\; i_{\vert I\vert}}=
\omega_{\nu i_1 i_2\dots\; i_{\vert I\vert}}}\eqno(5.4)$$
must hold for all ${i_1,\;\nu}$,
i.e. each
coefficient ${\omega_{i_1\nu i_2\dots i_{\vert I\vert}}}$\
is symmetric in the first two indices $i_1,\nu$. As a consequence
$\omega_{i_1\nu i_2\dots \; i_{\vert I\vert}}$\ resp.
$\omega_{i_1\nu i_1 i_3\dots \; i_{\vert I\vert}}$\
are also totally symmetric in the multi-index
$(i_3\dots i_{\vert I\vert})$\ for all $\vert I\vert$\ even
and greater than zero. This, however, contradicts the total antisymmetry
of the multi-index $I=(i_1,\dots\; i_{\vert I\vert})$.
Therefore
$\omega^i_{\mu i_1 \dots \; i_{\vert I\vert}}
=0$\  for all $2\le \vert I\vert\le 2n$\ and all $i$.
For $\vert I\vert=0$\ we have $c(dx^I):=\eins_{C(M)}$\ by definition
and therefore
obviously $\omega^i_{\mu}=0$. In sum we obtain only the
trivial solution  $\omega=0$. Consequently $\ker\;
{\it \Xi}=0$.
\medskip
$\underline{2.\ \rm case}$: $\End^+({\cal E})\cong \bigl(
C(M)^+\otimes
\End_{C(M)}^+({\cal E})\bigr)\;\oplus\;\bigl(
C(M)^-\otimes
\End_{C(M)}^-({\cal E})\bigr)$, i.e. the twisting graduation of
${\cal E}$\ is non-trivial.
\smallskip
We now only have to check the case of $\omega\in \Omega^1(M,C(M)^-\otimes
\End_{C(M)}^-
({\cal E}))$. Using $F_i\in \End_{C(M)}^-({\cal E})$, i.e.
$\vert F_i\vert=1$\ for all $i$\
together with the Clifford relation
we obtain from (3.8) as above
$${{\sum_{\vert I\vert\;\petit\rm odd}}\!^\prime\; \Bigl(
g^{\mu\nu}\omega^i_{\mu I}
c(dx^I) +(-1)^k g^{i_k\nu}\omega^i_{\mu I}
c(dx^\mu)c(dx^{i_1})\dots \widehat{ c(dx^{i_k})}\dots c(dx^{i_{\vert I\vert}}
)\Bigr) =0.}$$
For $\vert I\vert >1$\ we can argue as in the even case above
concluding
$\omega^i_{\mu i_1 \dots \; i_{\vert I\vert}}=0$\ for all $i$. In the
case $\vert I\vert=1$, however, we obtain
$${g^{\mu\nu}\omega^i_{\mu i_1} c(dx^{i_1})- g^{\nu i_1}\omega^i_{\mu
i_1}c(dx^\mu)=0.}$$
Therefore each coefficient
$\omega^i_{\mu\nu}=\omega^i_{\nu\mu}$\ is symmetric
and so $\omega=\omega^i_{\mu i_1}dx^\mu\!\otimes \!c(dx^{i_1}){\hat\otimes}
F_i\!\in Sym^2(T^*M){\hat\otimes}\End_{C(M)}^-({\cal E})$\ lies in
in the kernel of ${\it \Xi}$.
\QED
\smallskip
\Corollary There exist Dirac operators of simple type
not corresponding to Clifford connections on a Clifford module
${\cal E}$\ iff the twisting
graduation of ${\cal E}$\ is non-trivial. Moreover if
$\widetilde{D}$\ is such a Dirac operator, then
${\widetilde{D}=D_{\nabla^{\cal E}}+\eins_{C(M)}{\hat\otimes} \Phi,}$\
where $D_{\nabla^{\cal E}}$\ is a Dirac operator defined by a
Clifford connection $\nabla^{\cal E}$\ and $\Phi\in \Gamma(
\End^-_{C(M)}({\cal E}))$.
\smallskip
\Proof By the previous lemmas 5.2 and 5.3 the
first statement is obvious. So we only have to prove the second one.
\smallskip
Let $\omega:=\omega_{\mu\nu}^i dx^\mu\otimes c(dx^\nu){\hat\otimes}
F_i$\ with
$w^i:=\omega^i_{\mu\nu} dx^\mu\otimes c(dx^\nu)\in \Gamma(Sym^2(T^*M))\subset
\Gamma(T^*M\otimes C^-(M))$\ and $F_i\in \Gamma(\End_{C(M)}^-({\cal E})$\
be as in the previous
lemma. With the help of the Clifford relations we compute
$$\eqalign{c(\omega) &=\omega^i_{\mu\nu} c(dx^\mu) c(dx^\nu) {\hat\otimes}
F_i\cr
&={1\over 2}\;\omega^i_{\mu\nu}\bigl(c(dx^\mu)c(dx^\nu) + c(dx^\nu)
c(dx^\mu)\bigr){\hat\otimes} F_i\cr
&=- g^{\mu\nu} \omega_{\mu\nu}^i {\hat\otimes} F_i\cr
&=-tr(w^i){\hat\otimes} F_i\cr
&=:\eins_{C(M)}{\hat\otimes} \Phi \cr}\eqno(5.5)$$
with $\Phi:= \sum_i -tr(\omega^i)\cdot F_i\in \Gamma(
\End_{C(M)}^-({\cal E}))$.
Now let $\nabla^{\cal E}$\ be a Clifford connection and
let us define $\widetilde{\nabla}^{\cal E}:=\nabla^{\cal E}+\omega$.
We obtain $\widetilde{D}=
c(\widetilde{\nabla})=c(\nabla^{\cal E})+ c(\omega) =
D_{\nabla^{\cal E}}+ c(\omega)$\ for the corresponding Dirac operator
and the above computation
(5.5) proves the second statement.
\QED
\smallskip
Note that $\widetilde{\nabla}^{\cal E}$\ is not a Clifford connection
unless $\omega=0$. In the case of a twisted spinor bundle
${\cal E}= S\otimes E$\ the corresponding Clifford
superconnection $A$\ is given by
$${A:=\nabla^S\otimes \eins_E + \eins_S\otimes (\nabla^E+\Phi),}$$
i.e. $A$\ is uniquely determined by the
superconnection $A^E:=\nabla^E+\Phi$\ on the twisting bundle $E$.
\smallskip
We now study the decomposition formula (3.14) for the square of
such Dirac operators which are distinguished by lemma 5.3 resp. corollary
5.4 . Let
therefore again $\omega\in \Omega^1(M,\End^+({\cal E}))$\ be given by
$\omega:=w^i{\hat\otimes}
F_i$\ with
$w^i:=\omega^i_{\mu\nu} dx^\mu\otimes c(dx^\nu)\in \Gamma(Sym^2(T^*M))
\subset
\Gamma(T^*M\otimes C^-(M))$\ and $F_i\in
\Gamma(\End_{C(M)}^-({\cal E}))$.
Then we find that ${[c(dx^\mu), c(dx^\nu)][\omega_\mu,\omega_\nu]=0}$.
Thus, this term does not contribute to the endomorphism part
$F_{\widetilde{\nabla}^{\cal E}}$\ in (3.12).
Using this, straightforward computation yields
$${\widetilde{D}^2=\triangle^{\nabla^{\cal E}}
+{r_M\over 4} +
{\bf c}(R^{{\cal E}/S}_{\nabla^{\cal E}}) +
c\;\nabla^{{\petit\rm End}{\cal E}}(\eins_{C(M)}{\hat\otimes}\Phi) +
\eins_{C(M)}{\hat\otimes} \Phi^2}\eqno(5.6)$$
which can be seen as a
characteristic feature of Dirac operators of simple type. Now, if
one takes the connection laplacian ${\triangle^{\nabla^{\cal E}}}$\
together with
the Lichnerowicz part ${
{r_M\over 4} +
{\bf c}(R^{{\cal E}/S}_{\nabla^{\cal E}})}$,
then - by using (3.1) - we obtain the
\Corollary A Dirac operator $\widetilde{D}$\ acting on sections
of a Clifford module ${\cal E}$\ is of simple type
iff there exists a
Clifford connection $\nabla^{\cal E}$\ on ${\cal E}$\ and a
morphism $\Phi\in \Gamma(\End_{C(M)}^- {\cal E})$\ such that
${\widetilde{D}^2=D_{\nabla^{\cal E}}^2+
c\;\nabla^{{\petit\rm End}{\cal E}}(\eins_{C(M)}{\hat\otimes}\Phi) +
\eins_{C(M)}{\hat\otimes} \Phi^2}$.
\smallskip\rm
We note that the endomorphism part
in the decomposition formula
(5.6) of the square of a Dirac operator of simple type
is of Clifford degree $\le 2$. Getzler has recognized in [G], that this
is
the essential information needed to prove the local Atiyah-Singer
index theorem for Dirac operators associated with Clifford connections,
which is provided in this case by the `usual' Lichnerowicz formula
(3.1). As far as we know it has not yet been proven whether this
refined index theorem also holds for some Dirac operator not
associated to a Clifford connection. However, because of our observation
mentioned above and our formula (5.6) instead of
the usual Lichnerowicz formula (3.1), the techniques of
[BGV], chapter 4, can be adapted even to all
Dirac operators of simple type. So we state:
\Theorem For Dirac operators of simple type acting on sections of
a Clifford module ${\cal E}$\ the local Atiyah-Singer index theorem
holds.\rm

\vskip 1.0cm
{\mittel 6. The Wodzicki function}
\rm
\ukneu
\vskip 0.6cm
In this section, we define the Wodzicki function $W_{\cal E}$\ on the space
of all Dirac operators ${{\cal D}({\cal E})}$\ acting on sections
of a Clifford module ${\cal E}$\ via the non-commutative residue,
which has been studied extensively by Guillemin and Wodzicki (cf.
[Gu], [W1], [W2]). This function is closely related to
(gravity-) action functionals of physics as already mentioned
in the introduction and might also be useful to investigate
the space of all Dirac operators. This will be the subject of a
forthcoming paper. As will be seen, the
generalized Lichnerowicz
formula that we have derived in section 3 also applies to calculate
the Wodzicki function explicitly.
\medskip
Let $E$\ and $F$\ be finite dimensional complex vector bundles over
a compact, $n$-dimensional manifold $N$. The non-commutative
residue of
a pseudo-differential operator $P\in \Psi {\rm DO}(E,F)$\
can be defined by
$${res(P):={\Gamma({n\over 2})\over 2\pi^{n\over 2}}\;\int_{S^*M}\;
tr\bigl(\sigma^P_{-n}(x,\xi)\bigr)\;dx d\xi,}\eqno(6.1)$$
where $S^*M\subset T^*M$\ denotes the co-sphere bundle on $M$\ and
$\sigma^P_{-n}$\ is the component of order $-n$\ of the complete
symbol $\sigma^P:=\sum_i\;\sigma^P_i$\ of $P$. Here
the integral
is normalized by $vol(S^{n-1})^{-1}={\Gamma({n\over 2})\over 2
\pi^{n\over 2}}$.
In his thesis, Wodzicki has shown that the linear
functional $res\colon
\Psi {\rm DO}(E,F)\rightarrow \cz$\ is in fact
the unique trace (up to multiplication by constants) on
the algebra of pseudo-differential operators $\Psi{\rm DO}(E,F)$.
\smallskip
Now let $M$\ be a Riemannian manifold of even dimension $2n$\
and let
${\cal E}$\ be a Clifford module over $M$. If $D$\ is a Dirac operator
acting on sections of ${\cal E}$\ and $A\in \Gamma(\End^-({\cal E}))$,
then $\widetilde{D}:=D+A$\ is another Dirac operator corresponding to the
same Clifford action on ${\cal E}$. As we have already mentioned (cf.
section 2) the
converse is also true: Any two Dirac operators $D_0$\ and $D_1$\
differ by a section $A\in \Gamma(\End^-({\cal E}))$. Consequently
the
set of all Dirac operators ${\cal D}({\cal E})$\
corresponding to the same Clifford action
on a Clifford module
${\cal E}$\ is an affine space
modelled on $\Gamma(\End^-({\cal E}))$.
Hence, for any Dirac operator $D\in {\cal D}({\cal E})$\
we have the natural identifications
$${\Gamma(\End^-({\cal E}))\;\cong\;T_D{\cal D}({\cal E})\;\cong
{\cal D}({\cal E}).}\eqno(6.2)$$
\Definition Let ${\cal E}$\ be a Clifford module of rank $r$\
over a Riemannian manifold $M$\ with ${{\rm dim}\;M=2n}$\
and let ${\cal D}({\cal E})$\ be the space of all Dirac operators on
${\cal E}$.
The Wodzicki function $W_{\cal E}$\ on ${\cal D}({\cal E})$\
is the complex-valued function
$W_{\cal E}\colon {\cal D}({\cal E})
\rightarrow \cz$\ defined by $W_{\cal E}(D):=-{2\over
r(2n-1)}\;res(D^{-2n+2})$.
\smallskip\rm
Notice that in the case of fixing
a hermitian inner product $(\;\cdot\;,\;\cdot\;)_{\Gamma({\cal E})}$\ on
${\Gamma({\cal E})}$\ one can introduce the
(formal) adjoint operator $P^*$\
of $P\in \Psi {\rm DO}
({\cal E})$. Since $res(P^*)=\overline{res(P)}$\ (cf.[W2]), where
the bar denotes complex conjugation, the Wodzicki function
$W_{\cal E}$\ is real for self-adjoint Dirac operators.
\smallskip
As an intermediate
step to compute the Wodzicki function $W_{\cal E}$\ explicitly,
we need the following expression for the
diagonal part $\phi_1(x,x,\widetilde{D}^2)$\ of the subleading term
$\phi_1$\ in the heat-kernel expansion of the square
of an arbitrary Dirac operator $\widetilde{D}$:
\Lemma Let ${\cal E}$\ be a Clifford module over an even-dimensional
compact
Riemannian manifold $M$\ and let
$\widetilde{D}=c\circ\widetilde{\nabla}^{\cal E}$\ be
a Dirac operator. Then the diagonal part of the
subleading term $\phi_1$\ in the asymptotic expansion of
the heat-kernel of
$\widetilde{D}^2$\ is given by
$${\phi_1(x,x, \widetilde{D}^2)\!=\!{1\over 6} r_M(x)\!-\!\Bigl(\!
{\bf c}(R^{\widetilde{\nabla}^{\cal E}})\!
+\!ev_g\widetilde{\nabla}^{T^*M
\otimes {\petit \rm End}\;{\cal E}}\varpi_{\widetilde{\nabla}^{\cal E}}
\!+\!ev_g(\varpi_{\widetilde{\nabla}^{\cal E}}\cdot
\varpi_{\widetilde{\nabla}^{\cal E}})\!\Bigr)(x),}$$
with $\varpi_{\widetilde{\nabla}^{\cal E}}:=-{1\over 2} g_{\nu\kappa}
dx^\nu\otimes
c(dx^\mu)\bigl([\widetilde{\nabla}^{\cal E}_\mu, c(dx^\kappa)] +
c(dx^\sigma)\Gamma^\kappa_{\sigma\mu}\bigr)\in \Omega^1(M,
\End {\cal E})$.
\smallskip
\Proof Let $\hat\triangle$\ be a generalized laplacian acting on
sections of a hermitian vector bundle $E$\ over $M$\ and let
$\phi_1(x,x, {\hat\triangle})$\ denote the diagonal part of
the subleading term
$\phi_1$\ in the heat-kernel expansion of $\hat\triangle$. It is
well-known that
$${\phi_1(x,x,{\hat\triangle})={1\over 6}\;r_M(x)\cdot \eins_E-
F(x).}\eqno(6.3)$$
Again,
$r_M$\ denotes the scalar curvature of $M$\ and $F\in \Gamma(\End\;E)$\
is determined by the unique decomposition
of the generalized laplacian $\hat\triangle$:
$${{\hat\triangle}=\triangle^{{\hat\nabla}^E}+F.}\eqno(6.4)$$
Here
$\triangle^{{\hat\nabla}^E}$\ denotes the connection laplacian
associated with
the connection ${{\hat\nabla}^E}$\ on the hermitian bundle
${E}$.\par
In our case we have
$E={\cal E}$\ and ${{\hat\triangle}=\widetilde{D}^2}$. Using the
generalized Lichnerowicz formula (3.14) the further proof of
this lemma is obvious.
\QED
\smallskip
\Remark For an arbitrary generalized laplacian $\hat\triangle$\
on a hermitian vector bundle $E$\ it is only the
decomposition (6.4) which is proven to exist -
neither the
connection ${\hat\nabla}^E$\ nor the endomorphism $F\in \Gamma(\End\;E)$\
are known explicitly (cf. [BGV], Proposition 2.5). Consequently
in general
the subleading term $\phi_1(x,x, {\hat\triangle})$\ in the heat-kernel
expansion of ${\hat\triangle}$\ can not be computed with the help of
(6.4). Therefore lemma 6.2 is also interesting on its own.
\medskip
According to [KW]
the diagonal part $\phi_1(x,x,{\hat\triangle})$\ of the subleading
term in the heat-kernel expansion of
any generalized laplacian
$\hat\triangle$\ acting on sections of a hermitian vector bundle
$E$\ over an even dimensional Riemannian manifold $M$\
with ${{\rm dim}\;M=2n\ge 4}$\ can be related to
the non-commutative residue
$${res({\hat\triangle}^{-n+1})={2n-1\over 2}\;\int_M\;*tr\bigl(
\phi_1(x,x,{\hat\triangle})\bigr).}\eqno(6.6)$$
Here $*$\ denotes the Hodge-star operator defined by the
Riemannian metric on $M$. Together with the previous lemma we therefore
obtain:
\Theorem Let ${\cal E}$\ be a Clifford module over a compact,
Riemannian manifold $M$\ with ${{\rm dim}\;M=2n\ge 4}$\ and let
$\widetilde{D}=c\circ\widetilde{\nabla}^{\cal E}$\ be
a Dirac operator. Then
$${\!W_{\cal E}(\!\widetilde{D}\!)\!=\!\int_M *\Bigl(-{1\over 6}r_M
\!+\!{1\over r}
tr\bigl(
{\bf c}(R^{\widetilde{\nabla}^{\cal E}})\!+\!ev_g\widetilde{\nabla}^{T^*M
\otimes {\petit \rm End}\;{\cal E}}\varpi_{\widetilde{\nabla}^{\cal E}}
\!+\!ev_g(
\varpi_{\widetilde{\nabla}^{\cal E}}\cdot
\varpi_{\widetilde{\nabla}^{\cal E}})\bigr)\Bigr),}$$
with $\varpi_{\widetilde{\nabla}^{\cal E}}:=-{1\over 2} g_{\nu\kappa}
dx^\nu\otimes
c(dx^\mu)\bigl([\widetilde{\nabla}^{\cal E}_\mu, c(dx^\kappa)] +
c(dx^\sigma)\Gamma^\kappa_{\sigma\mu}\bigr)\in \Omega^1(M,
\End {\cal E})$.
\medskip\rm
Now let $({\cal E}_1, c_1))$\ and
$({\cal E}_2, c_2))$\ be Clifford modules of rank $r_1$\ resp. $r_2$\
over $M$. Here $c_i\colon
C(M)\times {\cal E}_i\rightarrow {\cal E}_i$\ for $i=1,2$\ denotes the
respective Clifford action of the Clifford bundle $C(M)$. Obviously
the direct sum bundle
${\cal E}:={\cal E}_1\oplus {\cal E}_2$\ together
with the Clifford action $c:=c_1\oplus c_2$\ is also a Clifford module.
Given Dirac operators
$\widetilde{D}_i:= c_i\circ \widetilde{\nabla}^{{\cal E}_i}$\
with $i=1,2$\ corresponding to the
connections $\widetilde{\nabla}^{{\cal E}_i}$\ on ${\cal E}_i$\ and
$a^{21}\in \Omega^1(M,\Hom({\cal E}_2,{\cal E}_1))$\ respectively
$a^{12}\in \Omega^1(M,\Hom({\cal E}_1,{\cal E}_2))$, the
following operator
$${\widetilde{D}:=\pmatrix{\widetilde{D}_1 &c_1\circ a^{21}\cr
                            c_2\circ a^{12}    &\widetilde{D}_2},
}\eqno(6.7)$$
is a
Dirac operator on ${\cal E}={\cal E}_1\oplus {\cal E}_2$.
\smallskip
We now want to calculate the Wodzicki function
$W_{\cal E}=W_{{\cal E}_1\oplus {\cal E}_2}$\ of this
Dirac operator $\widetilde{D}$. Given that $\widetilde{
\nabla}^{\cal E}:=\widetilde{\nabla}^{\cal E}\oplus
\widetilde{\nabla}^{{\cal E}_2}$\ be the direct sum connection,
an easy computation shows
$${\varpi_{\widetilde{\nabla}^{\cal E}}=\pmatrix{
\varpi_{\widetilde{\nabla}^{{\cal E}_1}} & b^{21}\cr
 b^{12} &\varpi_{\widetilde{\nabla}^{{\cal E}_2}}\cr}\in
\Omega^1(M,\End\;{\cal E}),}\eqno(6.8)$$
with $b^{ji}:= -{1\over 2} dx^\nu\otimes c_i(dx^\mu)\bigl(
a^{ji}_\mu c_j(dx^\nu)-c_i(dx^\nu)a^{ji}_\mu\bigr)\in
\Omega^1(M,\Hom({\cal E}_j, {\cal E}_i))$\ for $i,j\in \{1,2\}$\
and $i\ne j$. Denoting $a:=\bigl({0\atop a^{12}}{a^{21}\atop 0}\bigr)$\
and
$b:=\bigl({0\atop b^{12}}{b^{21}\atop 0}\bigr)$, so $a,b\in
\Omega^1(M,\End {\cal E})$, we further obtain:
$$\eqalign{tr\bigl(ev_g(\varpi_{\widetilde{\nabla}^{\cal E}}\cdot
\varpi_{\widetilde{\nabla}^{\cal E}})\bigr)=
tr\bigl(ev_g(\varpi_{\widetilde{\nabla}^{{\cal E}_1}}&\cdot
\varpi_{\widetilde{\nabla}^{{\cal E}_1}})\bigr) +
tr\bigl(ev_g(\varpi_{\widetilde{\nabla}^{{\cal E}_2}}\cdot
\varpi_{\widetilde{\nabla}^{{\cal E}_2}})\bigr) \cr
&+ tr\bigl(ev_g( b\cdot b)\bigr)
\cr}\eqno(6.9)$$
\smallskip
$$\eqalign{tr\bigl(ev_g(\widetilde{\nabla}^{T^*M\otimes
{\petit\rm End}\; {\cal E}}
\varpi_{\widetilde{\nabla}^{\cal E}})\bigr) =
tr\bigl(ev_g(\widetilde{\nabla}&^{T^*M\otimes {\petit\rm End}\; {
{\cal E}_1}}
\varpi_{\widetilde{\nabla}^{{\cal E}_1}})\bigr) \cr
&+
tr\bigl(ev_g(\widetilde{\nabla}^{T^*M\otimes {\petit\rm End}\; {
{\cal E}_2}}
\varpi_{\widetilde{\nabla}^{{\cal E}_2}})\bigr)\cr}\eqno(6.10)$$
and
$$\eqalign{tr\bigl({\bf c}(R^{\widetilde{\nabla}^{\cal E}})\bigr) =
tr\bigl({\bf c}_1(R&^{\widetilde{\nabla}^{{\cal E}_1}})\bigr) +
tr\bigl({\bf c}_2(R^{\widetilde{\nabla}^{{\cal E}_2}})\bigr) \cr
&+ tr\bigl({\bf c}([a\wedge a])\bigr).\cr}\eqno(6.11)$$
Here $[\;\cdot\;\wedge\;\cdot\;]$\ denotes the multiplication
in the graded Lie algebra $\Omega^*(M,\End {\cal E})$.\vfil\break
Thus we are able to prove the following
\Lemma Let ${({\cal E}_1, c_1)}$\ and ${({\cal E}_2,c_2)}$\ be
Clifford modules of rank $r_1$\ resp. $r_2$\ over a compact
Riemannian manifold $M$\ of even dimension
with ${2n\ge 4}$\ and let
$\widetilde{D}_1:=c_1\circ \widetilde{\nabla}^{{\cal E}_1}$\ resp.
$\widetilde{D}_2:=c_2\circ \widetilde{\nabla}^{{\cal E}_2}$\ be Dirac
operators. Then ${\widetilde{D}\!:=\!\bigl(\!{\!
\widetilde{D}_1 \atop c_2\circ a^{12}}\! {\! c_1\circ
a^{21}\atop
\widetilde{D}_2}\!\bigr)}$\ with $a:=\bigl(\!{0\atop a^{12}}\!
{a^{21}\atop 0}\!\bigr)\!\in\!\Omega^1(M,\End {\cal E})$\
defines a Dirac operator on ${\cal E}:=
{\cal E}_1\oplus {\cal E}_2$\ and
$${W_{\cal E}(\widetilde{D})={r_1\over r} W_{{\cal E}_1}(
\widetilde{D}_1)
\! +\!
{r_2\over r} W_{{\cal E}_2}(\widetilde{D}_2)\! +\!{1\over r}
\int_M\hskip -0.3cm *
tr\Bigr({\bf c}
([a\wedge a])\! +\!
ev_g
\Bigl(\hbox{${{b^{21}\cdot b^{12}
\atop 0}{0\atop b^{12}\cdot b^{21}}}$}\Bigr)
\Bigr)}$$
with
$b^{ji}:= -{1\over 2}g_{\kappa\nu} dx^\kappa\otimes c_i(dx^\mu)\bigl(
a^{ji}_\mu c_j(dx^\nu)-c_i(dx^\nu)a^{ji}_\mu\bigr)\in
\Omega^1(M,\Hom({\cal E}_j, {\cal E}_i))$\ for $i,j\in \{1,2\}$,
$i\ne j$\ and ${r:=(r_1+ r_2)}$.
\smallskip\rm
\vskip 1.0cm
{\mittel 7. The Wodzicki function for special Dirac operators}
\rm
\ukneu
\vskip 0.6cm
We now apply theorem 6.4 resp. lemma 6.5
to calculate the Wodzicki function of different types
of Dirac operators. These examples are inspired
by physics, where they might be used to derive gravity and
combined gravity/Yang-Mills actions, respectively
(cf. [K], [KW], [AT1], [AT2]). In the following, let
$M$\ be a
compact Riemannian manifold with ${{\rm dim}\;
M=2n\ge 4}$.
\vskip 0.4cm
{\bf $\bullet$ Dirac operators of simple type.} Let ${\cal E}$\ be
a Clifford module of rank $r$\ over
$M$\ and $\widetilde{D}$\ a Dirac operator of simple type
acting on sections of $\cal E$. By corollary 5.5 there exists
a Clifford connection $\nabla^{\cal E}$\ on $\cal E$\ and
an endomorphism $\Phi\in \End_{C(M)}^- ({\cal E})$, such that
we obtain the decomposition formula (5.6)
for the square of $\widetilde{D}$:
$${\widetilde{D}^2=\triangle^{\nabla^{\cal E}}
+{r_M\over 4} +
{\bf c}(R^{{\cal E}/S}_{\nabla^{\cal E}}) +
c\;\nabla^{{\petit\rm End}{\cal E}}(\eins_{C(M)}{\hat\otimes}\Phi) +
\eins_{C(M)}{\hat\otimes} \Phi^2.}$$
Using theorem 6.4 we then calculate
$$\eqalign{\!\!W_{\cal E}\!(\!\widetilde{D}\!)\!&=\!\int_M\!*\!\Bigl(\! -
{1\over 6}r_M\! +\!{1\over r}
 tr\bigl(
{r_M\over 4}\eins_{\cal E}\! +\!
{\bf c}(\!R^{{\cal E}/S}_{\nabla^{\cal E}}\!) \!+\!
c\!\nabla^{{\petit\rm End}{\cal E}}(\!\eins_{C(M)}\!{\hat\otimes}\!\Phi\!
)\! +\!
\eins_{C(M)}\!{\hat\otimes}\! \Phi^2\!\bigr)\Bigr)\cr
&=\int_M\;*\Bigl({1\over 12}\;r_M +{2^n\over r}\;tr(\Phi^2)\Bigr)\cr
}$$
for the Wodzicki function,
since both $tr\bigl({\bf c}(R^{{\cal E}/S}_{\nabla^{\cal E}})\bigr)$\
and $tr\bigl(c\nabla^{{\petit\rm End}\;{\cal E}}(\eins_{C(M)}{\hat
\otimes}\Phi)\bigr)$\ vanish. \smallskip
Now let us take the zero-morphism for
$\Phi$. In that exceptional case
$\widetilde{D}$\ is associated with
the Clifford connection $\nabla^{\cal E}$. Therefore we obtain
 $W_{\cal E}(\widetilde{D})={1\over 12}
\int_M *r_M$. Thus, the Wodzicki function evaluated on Dirac operators
corresponding to Clifford connections reproduce the classical
Einstein-Hilbert functional on $M$. This was already mentioned by Connes
and explicitly shown in [K] and [KW].
In the case of $M$\ being a spin-manifold and
${\cal E}=S\otimes E$\ with $E:=M\times \cz^2=(M\times\cz)\oplus
(M\times \cz)$, then $\End{\cal E}=C(M){\hat\otimes}M_2(\cz)$. Here
$M_2(\cz)$\ denotes the algebra of two by two matrices over
the complex numbers. Interesting, we thus recover the situation of a
`non-commutative
two-point space' as considered in [CFF] and [KW]. Further
specializing
the Dirac operator of simple type
by
$\Phi:=\phi\cdot \bigl({0\atop 1}{1\atop 0}\bigr)$, where
$\phi\in C^\infty(M)$\ denotes
a complex-vallued function as in the mentioned references,
we obtain
$${W_{\cal E}(\widetilde{D})=\int_M *\bigl({1\over 12}r_M+\phi^2\bigr).}\eqno(
7.1)$$
This was interpreted in [KW] as Einstein-Hilbert gravity action with
cosmological constant.
\vskip 0.4cm
{\bf $\bullet$ Dirac operators with torsion.}
Let now $M$\ be a spin manifold in addition.
The Levi-Civita connection ${\nabla\colon \Gamma(TM)\rightarrow
\Gamma(T^*M\otimes TM)}$\ on $M$\ induces a
a connection ${\nabla^S\colon
\Gamma(S)\rightarrow \Gamma(T^*M\otimes S)}$\ on the spinor bundle
$S$\ which is compatible with the hermitian metric
$<\;\cdot\;,\;\cdot\;>_S$\ on $S$. Adding a
torsion term $t\in \Omega^1(M,\End TM)$\ to the Levi-Civita connection,
we obtain a new covariant derivative
${\widetilde{\nabla}:=\nabla + t}$\
on the tangent bundle $TM$. Since $t$\ is really
a one-form on $M$\ with values in the bundle of skew endomorphism
$Sk(TM)$\ (cf. [GHV]), $\widetilde{\nabla}$\ is in fact
compatible with the Riemannian metric $g$\ and
therefore it also induces a connection
$\widetilde{\nabla}^S=\nabla^S+T$\ on the
spinor bundle. Here $T\in \Omega^1(M,\End S)$\ denotes the
`lifted' torsion term $t\in \Omega^1(M,\End TM)$.
However, in general this induced connection $\widetilde{\nabla}^S$\
is neither compatible with the hermitian metric $<\;\cdot\;,\;\cdot\;
>_S$\ nor is it a Clifford connection.
\smallskip
The
most general
Dirac operator on the spinor bundle $S$\ corresponding to
a metric connection $\widetilde{\nabla}$\ on $TM$\ can be defined by
${\widetilde{D}:=c\circ\widetilde{\nabla}^S}$. Thus
the Wodzicki function $W_{\cal E}$\ of $\widetilde{D}$\
yields
$${W_{\cal E}(\widetilde{D})=\int_M *\Bigl(
{1\over 12}\;r_M
+{1\over 2^n}\; \bigl(- t_{abc}t^{abc}+2t_{abc}t^{acb}\bigr)\Bigr).
}\eqno(7.2)$$
 From a physical point of view, this can be interpreted as
the action functional
of a modified Einstein-Cartan theory (cf. [AT1]).
\vskip 0.4cm
{\bf $\bullet$ An Einstein-Yang-Mills Dirac operator.} Let
$({\cal E},c)$\ be
a Clifford module of rank $r= 2^n\cdot rk({\cal E}/S)$, where
${\cal E}/S$\ denotes the twisting part of ${\cal E}$. As already
mentioned, the direct sum ${{\bar{\cal E}}:={\cal E}\oplus {\cal E}}$\
is also a Clifford module. Given a Clifford connection
$\nabla^{\cal E}$\ on ${\cal E}$, obviously
$\nabla^{\bar{\cal E}}:=\bigl({\nabla^{\cal E}\atop 0}{0\atop
\nabla^{\cal E}}\bigr)$\ defines a Clifford connection on
${\bar{\cal E}}$. Define
$${\widetilde{\nabla}^{\bar{\cal E}}:=\nabla^{\bar{\cal E}} +
\Bigl(\hbox{${{0\atop -a}{a\atop 0}}$}\Bigr),}\eqno(7.3)$$
with
$a:=dx^\mu\otimes c(dx^\nu){\hat\otimes}
R_{\mu\nu}\!\in\! \Omega^1(M,\End{\cal E})$. Here
$R=R^{{\cal E}/S}_{\nabla^{\cal E}}\!\in\! \Omega^2(M,
\End_{C(M)}({\cal E}))$\ denotes the
twisting curvature of $\nabla^{\cal E}$. We now consider the
associated Dirac operator
$${\widetilde{D}:={\bar c}\circ \widetilde{\nabla}^{\bar {\cal E}} =
\pmatrix{D_{\nabla^{\cal E}} & c\circ a\cr
          c\circ (-a) &D_{\nabla^{\cal E}} \cr}.}\eqno(7.4)$$
An easy computation yields $\varpi_{\nabla^{\bar{\cal E}}}=
\bigl({0\atop -a}{a\atop 0}\bigr)$. Thus, with regard to our
lemma 6.5 we obtain
$$\eqalign{tr\; {\bar {\bf c}}\Bigl(\Bigl[\bigl(\hbox{${{0\atop -a}
{a\atop 0}}$}\bigr)
\wedge \bigl(\hbox{${{0\atop -a}{a\atop 0}}$}\bigr)\Bigr]\Bigr) &=
-2^n\; tr \Bigl(\hbox{${{R_{\mu\nu}
R^{\mu\nu}\atop 0}{0\atop
R_{\mu\nu}R^{\mu\nu}}}$}\Bigr)\cr
\noalign{\vskip 0.1cm}
tr\; ev_g \Bigl(\hbox{${{-a\cdot a\atop 0}{0\atop -a\cdot a}}$}\Bigr) &=
-2^n\; tr \Bigl(\hbox{${{R_{\mu\nu}R^{\mu
\nu}
\atop 0}{0\atop
R_{\mu\nu}R^{\mu\nu}}}$}
\Bigr).\cr}\eqno(7.5)$$
When we use lemma 6.5 together with our first example, the
Wodzicki function $W_{\bar{\cal E}}$\ for the above defined
Dirac operator $\widetilde{D}$\ yields
$${W_{\bar {\cal E}}(\widetilde{D}) =
=\int_M *\Bigl({1\over 12 }\;r_M -{2\over rk({\cal E}/S)}
\;
tr(R_{\mu\nu}R^{\mu\nu})\bigr).
}\eqno(7.6)$$\nobreak
Thus we recover the combined Einstein-Hilbert/Yang-Mills action. Note,
that this
example can also be understood in the sense of a non-commutative
two-point space (compare the first example).
\smallskip
Although being a gauge theory, it is well-known that
the classical theory of
gravity as enunciated by Einstein stands apart from the non-abelian
gauge field theory of Yang and Mills, which encompasses the three
other fundamental forces: the weak, strong and electromagnetic
interactions. Interesting, as this example shows, they both have
a common `root': the special Dirac operator $\widetilde{D}$\
considered above. We hope that this may shed new light on the
gauge theories in question. Moreover, in a physical sense the above
derivation of (7.6) via the Wodzicki function $W_{\bar {\cal E}}$\
can be understood
as unification of Einstein's gravity- and Yang-Mills theory as it is shown
in [AT2].
\vskip 1cm
{\mittel 8. Conclusion}
\vskip 0.6cm
In this paper we studied Dirac operators acting on sections of a
Clifford module ${\cal E}$\ over a Riemannian manifold $M$. We obtained
an intrinsic decomposition of their squares, which is the
generalisation of the well-known Lichnerowicz formula [L]. This
enabled us to distinguish Dirac operators of simple type. For each
operator of this natural class the local Atiyah-Singer index theorem
was shown to hold. Furthermore we defined a complex-vallued
function $W_{\cal E}\colon {\cal D}({\cal E})\rightarrow \cz$\ on the
space of all Dirac operators on ${\cal E}$, the Wodzicki function,
via the non-commutative
residue. If $M$\ is compact and ${{\rm dim}\; M=2n\ge 4}$,
we derived an expression for $W_{\cal E}$\ in theorem 6.4 . For
certain Dirac operators we calculated this function explicitly. From
a physical point of view, this
provides a method to reproduce gravity, resp. combined gravity/Yang-Mills
actions out of the Dirac operators in question. Therefore we
expect new insights in the interrelation of Einstein's gravity
and Yang-Mills gauge theories.
\vskip 0.5cm\nobreak
{\bf Acknowledgements.} We would like to thank E.Binz for his gentle
support and Susanne for carefully reading the manuscript.
\goodbreak
\begref
\ref{[AT1]} T.Ackermann, J.Tolksdorf,  \sl
A generalized Lichnerowicz formula, the Wodzicki residue and Gravity\rm ,
Preprint CPT-94/P.3106 and Mannheimer Manuskripte 181, (1994)
\ref{[AT2]} T.Ackermann, J.Tolksdorf, \sl
Dirac operators, Einstein's gravity and Yang-Mills theory\rm , to appear
\ref{[AS]} M.F.Atiyah, I.M.Singer, \sl
The index of elliptic operators I\& III, Ann. of Math. {\bf 87}
(1968) 484-530 resp. 546-604
\ref{[CFF]} A.H.Chamseddine, G.Felder, J.Fr\"ohlich, \sl
Gravity in non-commutative geometry\rm , Com. Math. Phys {\bf 155} (1993),
205-217
\ref{[BGV]} N.Berline, E.Getzler, M.Vergne, \sl Heat kernels and
Dirac operators\rm , Springer (1992)
\ref{[C]} A.Connes, \sl Non-commutative geometry and physics\rm ,
IHES preprint (1993)
\ref{[CL]} A.Connes, J.Lott,\sl Particle models and non-commutative
geometry,\rm Nucl. Phys. B Proc. Supp. {\bf 18B} (1990) 29-47
\ref{[G]} E.Getzler, \sl A short proof of the local Atiyah-Singer
index theorem\rm , Topology {\bf 25} (1986), 111-117
\ref{[GHV]} W.Greub, S.Halperin, R.Vanstone, \sl Connections, curvature
and cohomology\rm , vol. 1, Academic press (1976)
\ref{[Gu]} V.Guillemin, \sl A new proof of Weyl's formula on the
asymptotic distribution of eigenvalues\rm , Adv. Math. {\bf 55} (1985),
131-160
\ref{[L]} A.Lichnerowicz, \sl Spineurs harmonique\rm ,
C. R. Acad. Sci. Paris S${\acute {\rm e}}$r. A {\bf 257} (1963)
\ref{[K]} D.Kastler, \sl The Dirac operator and gravitation\rm ,
to appear in Com. Math. Phys.
\ref{[KW]} W.Kalau, M.Walze, \sl Gravity, non-commutative geometry
and the Wodzicki residue\rm , to appear in Journ. of
Geometry and Physics
\ref{[W]} E.Witten, \sl A new proof of the positive energy theorem\rm ,
Com. Math. Phys. {\bf 80} (1981), 381-402
\ref{[W1]} M.Wodzicki, \sl Local invariants of spectral asymmetry\rm ,
Inv. Math. {\bf 75}, 143-178
\ref{[W2]} M.Wodzicki, \sl Non-commutative residue I\rm , LNM {\bf 1289}
(1987), 320-399
\bye